\documentclass[aps,pra,twocolumn]{revtex4}
\usepackage{slashed}
\usepackage{amsfonts}
\usepackage{bbding}
\usepackage{amssymb}
\usepackage{color}
\usepackage{amsmath}
\usepackage{graphicx}
\usepackage{subfigure}
\usepackage{txfonts}
\usepackage{bm}
\usepackage{ulem}
\usepackage[colorlinks,linkcolor=blue,citecolor=blue]{hyperref}
\newcommand{\jj}[1]{{\color{red}#1}}

\begin{document}

\title{Synchronization of persistent oscillations in spin systems with non-local dissipations}
\author{Xingli Li\footnote{Present address: Department of Physics, The Chinese University of Hong Kong, Shatin, New Territories, Hong Kong, China}, Yan Li, and Jiasen Jin}
\email{jsjin@dlut.edu.cn}
\affiliation{School of Physics, Dalian University of Technology, 116024 Dalian, China}
\date{\today}
\begin{abstract}
We explore the synchronization phenomenon in the quantum few-body system of spins with the non-local dissipation. Without the external driving, we find that the system can exhibit stable oscillatory behaviors in the long-time dynamics accompanied by the appearance of the purely imaginary eigenvalues of the Liouvillian. Moreover, the oscillations of the next-nearest-neighboring spins are completely synchronized revealed by the quantum trajectory analysis within the stochastic Schr{\"o}dinger equation. The possibility of the appearance of the long-time oscillations in infinite-size lattice by means of cluster mean-field approximation is also discussed.
\end{abstract}

\maketitle
\section{Introduction}
\label{Introduction}
Classical synchronization is a multi-disciplinary and fascinating topic that is found in abundance in both the natural and social sciences, e.g., applause, traffic lights, heart cells, etc.\cite{BookSYNC,CaiPRL1993,RosinPRL2013}. The phenomenon of synchronization was first noticed by Huygens in the 17th century \cite{HuygensSYNC} and classical synchronization has been extensively studied in various fields such as physics, chemistry and biology over the past decades \cite{Arenas2008,Manrubia2004,BoccalettiRP2002,Strogatz2001}. In general, the synchronization can be classified into two different types according to its generation mechanism: forced synchronization \cite{GoychukPRL2006,GalvePRL2010,ZhirovPRL2008,ZhirovPRB2009} and spontaneous synchronization \cite{WalterPRL2014,AmeriPRA2015}. In contrast to the forced case, the spontaneous synchronization is not driven by any external driving force but only as a consequence of the interaction between the subsystems that sharing the similar time-dependent properties.

Quantum synchronization, as the extension of classical synchronization in the quantum regime, has also received considerable attention in past few decades. The studies on quantum synchronization, on the one hand, focus on quantifying the degree of quantum synchronization. Unlike those technologies that have been well-established and widely used in the quantification of classical synchronization, the absence of the clear notion of the phase-space trajectories prevents the straightforward extension of classical synchronization measures into the quantum regime \cite{GalvearXiv,Mari2013PRL}. Additionally, one of the most significant difference between quantum and classical systems is that there may be non-local correlations between subsystems in quantum systems. Such correlations may play significant role in the time-evolution of the subsystems and in turn is a key factor for the quantum synchronization. Thus it is necessary to take into account the quantum correlations in defining the measures of quantum synchronization \cite{GalvearXiv}. For this purpose, measures of quantum synchronization through either quantum correlations between local operators, e.g., synchronization error \cite{Mari2013PRL}, correlations of observables \cite{zhuNJP2015}, or information-based quantum correlations, e.g., mutual information \cite{GiorgiPRA2012}, entanglement \cite{LEEPRE2014} are proposed. Until now, certain desirable methods have already been designed and developed for different systems and conditions to quantify the quantum synchronization \cite{Jaseem2020PRE,Karpat2020PRA,Mari2013PRL,GiorgiPRA2012,LEEPRE2014,zhuNJP2015,GalvearXiv,AmeriPRA2015,OrthPRB2010,LeePRA2013,Jaseem2020PRR,Solanki2022PRAL}.

Looking for the quantum systems which may exhibit the phenomenon of synchronization is another fascinating investigation direction of quantum synchronization. Prior to this, quantum synchronization has already been explored in plenty of different open quantum systems\cite{Krithika2022PRA,Laskar2020PRL,GalvearXiv}, such as the van der Pol oscillators \cite{Walter2014PRL,Lee2013PRL,Tilley2018NJP}, atomic ensembles \cite{Xu2014PRL}, superconducting circuit systems \cite{Vinokur2008Nature,Quijan2013PRL} and optomechanical systems \cite{Ludwig2013PRL,Cabot2017NJP}. Plenty of the theoretical and experimental explorations have already been carried out \cite{Agrawal2013PRL,Matheny2014PRL,Zhang2012PRL}. Moreover, the relevant studies have even been explored in the so-called collision model which is another microscopic description of open quantum system. In the collision model framework, Karpat {\it et. al.} proposed a scheme with particle free evolutionary processes to study the environment induced spontaneous synchronization \cite{Karpat2019PRA}. Basing on the setting of stroboscopic collisions, the mutual synchronization between two spin-1/2 particles, characterized by the Person correlation coefficient, is established. The work also provides a novel path to explore the continuous oscillations in open quantum systems.

Usually the dynamics of the open quantum system governed by the quantum master equation in Lindblad form will always point to a time-independent steady state. This characterization, as stated by Evans's theorem \cite{Evans1979}, is determined by the structure of the corresponding Liouvillian \cite{Albert2014PRA,Albert2016PRX,Cattaneo2020PRA,Buca2022Sci}. It is worthy to notice that apart from the asymptotic steady states, the non-stationary state can appear in some specific dissipative quantum systems, such as the driven-dissipative spin systems \cite{Chan2015PRA,TurnerNP2018} and periodically driven systems \cite{Else2016PRL}.

In particular, to the best of our knowledge, Bu\v ca {\it et. al.} argued for the first time that if there exists an eigenoperator of the Hamiltonian that commutes with all the jump operators, the long-time oscillation will be established in quantum many-body systems \cite{CsmunozPRA2019, JakschNC2019,TindallNJP2020}. Such oscillations are directly induced by the dark states of the so-called dark Hamiltonian. Therefore, under this condition, the dissipation splits the whole state space into decaying and non-decaying state spaces. The latter can also be considered as the dynamical decoherence-free space. Due to the dissipation induced by the external environment, the states in the decaying state space will eventually disappear like being erased during the evolution.  However the dark states can be preserved during the dissipative process and they construct the different disjoint sectors of dynamical decoherence-free space. In this case, the system can oscillate between disjoint sectors under the driven of the dark Hamiltonian \cite{CsmunozPRA2019, JakschNC2019,TindallNJP2020}.

In this paper, we introduce the non-local dissipations into a four-body spin-1/2 system on a square lattice. The next-nearest neighboring spins do not couple to each other via direct interactions but through sharing the same environment. With the help of the environment-induced driving we find that the present system evolves to a non-stationarity oscillator in the long-time limit ($t\rightarrow\infty$). Moreover we show that the oscillations for each subsystems are synchronized.

We start our discussions by considering two types of Hamiltonian: the spin-1/2 XXZ and XYZ models. After showing the spectrum of the Liouvillian in the complex plane, we confirm the existence of the purely imaginary eigenvalues in the XXZ model which implies that the sustaining oscillations may be found in this system. We perform the fast Fourier transform (FFT) to numerically determine the dominant frequency of the oscillation, then we study the synchronization properties via the stochastic quantum trajectories. In order to investigate the robustness of the oscillations to small perturbations, we calculate the largest Lyapunov exponent \cite{YusipovChaos2019,XingliPRB2021,YusipovChaos2021,YusipovChaos2022} through the single quantum trajectory and check the stability of oscillation. We discuss the possibility of the existence of oscillations in the infinite-size system with the help of the cluster mean-field approximation \cite{JinPRX2016} in the last part of the paper.

The paper is organized as follows. In Sec.\ref{Model and Method} we start  with a brief introduction of the considered models and review the mechanism that induces the non-stationarity state. We present the definitions of the quantum synchronization measures as the preparation for further discussions. In Sec.\ref{results}, we discuss the dynamical properties of the systems in details. The Liouvillian spectra are first investigated to provide a general idea of the dynamical properties of the systems. We then discuss the sublattice symmetry of the (oscillating) steady state and uncover the dark state subspace that the oscillation lies in. We focus on the dissipative XXZ model to verify the existence of the persistent oscillation of the observable in the realistic time-evolution. We also investigate the time-dependence of two quantities, Loschmidt echo and the purity, those concerned in experimental detection. The stability of the oscillations in the spin dynamics is discussed via the largest Lyapunov exponent. The synchronization between the spins are discussed by means of the single quantum trajectories in the stochastic Schr\"odinger equation (SSE). We also investigate the sublattice symmetry of the (oscillating) steady state. The possibility of the appearance of the oscillatory phase in the infinite-size system is also discussed with the cluster mean-field method. We summarize in Sec.\ref{Summary}.

\section{Models and Methods}
\label{Model and Method}

We consider a dissipative system with four spin-1/2 subsystems on a square lattice and their interaction is the general XYZ exchange Hamiltonian (hereinafter $\hbar=1$),
\begin{equation}
\hat{H} = \sum_{\langle j,l\rangle}(J_{x}\hat{\sigma}^{x}_{j}\hat{\sigma}^{x}_{l} + J_{y}\hat{\sigma}^{y}_{j}\hat{\sigma}^{y}_{l} + J_{z}\hat{\sigma}^{z}_{j}\hat{\sigma}^{z}_{l}),
\label{Hamiltonian}
\end{equation}
where $\hat{\sigma}^{\alpha}_{j}(\alpha = x,y,z)$ are the Pauli matrices for the $j$-th spin and $\langle j,l\rangle$ indicates the summation of the interactions runs over the nearest-neighboring sites. The XYZ exchange Hamiltonian is generic in the spin systems and can be reduced to the XXZ model Hamiltonian for the case with $J_{x}=J_{y}\neq J_{z}$.

\begin{figure}[!htpb]
\includegraphics[width=0.3\textwidth]{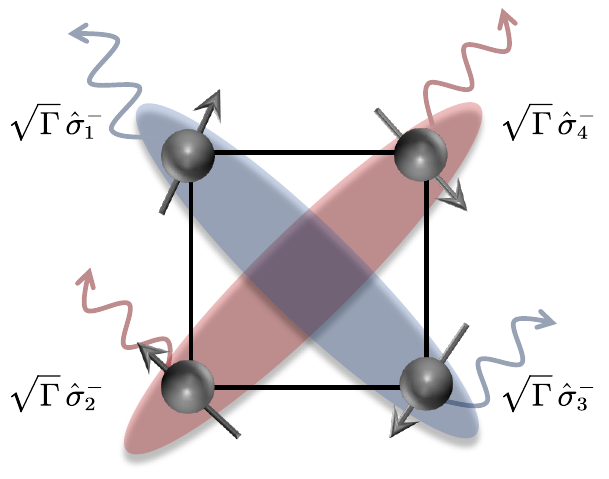}
\caption{\label{Schematic} Schematic illustration of the considered spin model. Each spin directly interacts with its nearest neighbor via the XYZ or XXZ interactions. In addition, each spin together with its next-nearest neighbor simultaneously couples to the same bath resulting in a non-local dissipation. }
\end{figure}

As shown in Fig.\ref{Schematic}, the composite system interacts with the environments and is consequently driven away from equilibrium. Moreover, the next-nearest neighboring spins are considered to couple with the same bath. Thus the dynamics of the system in the presence of the non-local dissipations can be described by the following Lindblad master equation,
\begin{equation}
\frac{\partial\hat{\rho}(t)}{\partial t} = \mathcal{L}[\hat{\rho}(t)]= -i[\hat{H},\hat{\rho}(t)] + \mathcal{D}[\hat{\rho}(t)],
\label{Masterequation}
\end{equation}
where $\mathcal{L}[\cdot]$ is the Liouvillian superoperator and the non-local dissipator $\mathcal{D}[\cdot]$ is given by
\begin{equation}
\mathcal{D}[\hat{\rho}(t)] = \sum_{k=1,2}\mathcal{\hat{O}}_{k}\hat{\rho}(t)\mathcal{\hat{O}}^{\dagger}_{k} - \mathcal{\hat{O}}^{\dagger}_{k}\mathcal{\hat{O}}_{k}\hat{\rho}(t)/2 - \hat{\rho}(t)\mathcal{\hat{O}}^{\dagger}_{k}\mathcal{\hat{O}}_{k}/2.
\label{dissipator}
\end{equation}
The non-local dissipations are generated by the joint jump operators $\mathcal{\hat{O}}_{1} = \sqrt{\Gamma}(\hat{\sigma}^{-}_{1} + \hat{\sigma}^{-}_{3})$ and $\mathcal{\hat{O}}_{2} = \sqrt{\Gamma}(\hat{\sigma}^{-}_{2} + \hat{\sigma}^{-}_{4})$ with $\Gamma$ being the decay rate. The operator $\hat{\sigma}^{-}_{j} = (\hat{\sigma}^{x}_{j} - i\hat{\sigma}^{y}_{j})/2$ is the local lowering operator for the $j$-th spin and describes the incoherent spin-flip process that tends to align the spin down to the $z$ direction. We will work in units of $\Gamma$ in the rest of this paper.

The nonlocal dissipation discussed here is an intermediate case of the commonly considered scenarios, the independent decay and collective decay. The nonlocal dissipation has recently been discussed for the correlation engineering \cite{seetharam2022prr,seetharam2022prb}, phase transitions \cite{parmee2018} and time crystals in dissipative systems \cite{lledo2019,passarelli2022}. It can be realized by coupling the two involving spins to a common lossy cavity mode. The detection of a leaked photon does not allow one to distinguish which spin emitted the photon \cite{seetharam2022prr}. Besides, an external plane wave illuminating on the two spins can also induced the nonlocal dissipations and is tunable via the spacing of the spins \cite{parmee2018}. The schemes are feasible with the state-of-the-art platform of atomic quantum simulator \cite{norcia2018,vaidya2018,bentsen2019}.

The eigensystem of the Liouvillian superoperator in Eq. (\ref{Masterequation}) reads
\begin{equation}
\mathcal{L}[\hat{\rho}_{j}] = \lambda_{j}\hat{\rho}_{j},
\label{spectrum}
\end{equation}
\begin{equation}
\mathcal{L}^\dagger[\hat{\eta}_{j}] = \lambda_{j}^*\hat{\eta}_{j},
\label{spectrum_conj}
\end{equation}
where $\lambda_j$ ($j=0,1,2,...$) are the complex eigenvalues of $\mathcal{L}$. The right and left eigenoperators, $\hat{\rho}_{j}$, $\hat{\eta}_{j}$, satisfies the following bi-orthogonal relation,
\begin{equation}
\langle\langle\hat{\eta}_j|\hat{\rho}_{j'}\rangle\rangle=\delta_{j,j'},
\end{equation}
where $\langle\langle\hat{\eta}_j|\hat{\rho}_{j'}\rangle\rangle=\text{tr}(\hat{\eta}^\dagger_j\hat{\rho}_{j'})$ is the Hilbert-Schmidt inner product.
Note that the superoperator $\mathcal{L}$ generates a completely positive and trace-preserving dynamical map that describes a physical time-evolution, thus the real parts of these eigenvalues are always semi-negative definite, i.e. $\text{Re}[\lambda_j]\le0$, $\forall j$. We sort the eigenvalues by the real parts in descending order as $0\ge\text{Re}[\lambda_{0}]\ge\text{Re}[\lambda_{1}]\ge\text{Re}[\lambda_{2}]\ge\cdots$.  There exists at least one eigenvalue whose real part is zero.

Combine Eqs. (\ref{Masterequation}) and (\ref{spectrum}), the time-evolution of the state of the system can be formally obtained as \cite{KesslerPRA2012,MingantiPRA2018},
\begin{equation}
\hat{\rho}(t)=e^{\mathcal{L}t}\hat{\rho}(0)\sim\hat{\rho}_{\text{ss}} + \sum_{j\neq0}c_{j}e^{\lambda_{j}t}\hat{\rho}_{j},
\label{time-evolution equation}
\end{equation}
where $c_{j}$ is the probability amplitude of the initial state $\hat{\rho}(0)$ in the basis of
the eigenstates of the Liouvillian. In the long-time limit, the system will eventually reach an asymptotic steady state $\hat{\rho}_{\text{ss}}=\lim_{t\to\infty}e^{\mathcal{L}t}\hat{\rho}(0)$. Actually, the steady state corresponds to the eigenstate $\hat{\rho}_{0}$ associated to the eigenvalue $\lambda_0$ that satisfying $\text{Re}[\lambda_{0}] = 0$, namely $\hat{\rho}_{\text{ss}} = \hat{\rho}_{0}/\text{Tr}[\hat{\rho}_{0}]$. In addition, the largest nonzero real part of the eigenvalues is also defined as the asymptotic decay rate which determines the slowest relaxation in the dynamics \cite{KesslerPRA2012,MingantiPRA2018}.

Generally, there is always a stationary steady-state solution to the master equation (\ref{Masterequation}) which satisfies the formula $\mathcal{L}[\hat{\rho}_{\text{ss}}]=0$. However, according to the specific form of Eq. (\ref{time-evolution equation}), we can find that when there are purely imaginary eigenvalues of the Liouvillian,  the system has non-stationary steady state manifested by an oscillation in the long-time limit. In the viewpoint of the Liouvillian, Eq.(\ref{spectrum}) can be rewritten as $\mathcal{L}[\hat{\rho}_{j}] = -i\mathcal{H}[\hat{\rho}_{j}]=\lambda_j\hat{\rho}_{j}$ where $\mathcal{H}$ is the superoperator associated to the dark Hamiltonian. When there exists a subset of states $\{|\psi_{1}\rangle,|\psi_{2}\rangle...,|\psi_{j}\rangle\}$ where $|\psi_{j}\rangle$, $\forall j$, are the eigenstates of the system Hamiltonian but are the dark states of any of the jump operators $\mathcal{\hat{O}}_{k}$, i.e. $\mathcal{\hat{O}}_{k}|\psi_{j}\rangle=0$. We call all the $|\psi_j\rangle$ the dark states of $\mathcal{H}$. The set of dark states can span a subspace in which one can define the so-called pseudo-density matrices satisfying $\mathcal{L}[|\psi_{j}\rangle\langle\psi_{l}|] = i(\omega_{l}-\omega_{j})|\psi_{j}\rangle\langle\psi_{l}|$. It is easy to find that such pseudo-density matrices exhibit the oscillatory behaviors during the time-evolution.

In fact, the pseudo-density matrices can be used to construct a valid density matrix whose diagonal elements are time-independent while the off-diagonal elements are time-dependent. Thus, governed by the dark Hamiltonian, the dynamics can be considered as a state-erasing process because only the dark states are survived at the end of the evolution. Then the subspace is decomposed into different disjoint sectors depending on the eigenvalues. The state of the system varies in different sectors resulting in a time-dependent steady state. Note that the properties of the oscillatory state are determined by the initial state and structures of the subspaces \cite{CsmunozPRA2019,JakschNC2019,TindallNJP2020}.

As the long-time oscillations are established, it is interesting to investigate the degree of synchronization among the spins.
Following the proposal in Ref. \cite{NajmehPRR2020}, we employ the following temporal complex-valued correlator as a measure of the synchronization,
\begin{equation}
C_{jl}(t)= |C_{jl}|e^{i\phi_{jl}t}=\frac{\langle \hat{\sigma}^{+}_{j}\hat{\sigma}^{-}_{l}\rangle_{t}}{\sqrt{\langle \hat{\sigma}^{+}_{j}\hat{\sigma}^{-}_{j}\rangle_{t}\langle \hat{\sigma}^{+}_{l}\hat{\sigma}^{-}_{l}\rangle_{t}}}.
\label{Cij}
\end{equation}
The phase $\phi_{jl}\in[-\pi,\pi]$ characterizes the phase difference between the $j$-th and $l$-th spins and $\langle\hat{o}\rangle_{t}=\text{Tr}[\hat{o}\hat{\rho}(t)]$ denotes the expectation value of a given operator $\hat{o}$ at the time $t$.
The correlator $C_{jl}(t)$ was originally proposed to measure the phase-locking and synchronization of two quantum harmonic oscillators. When the oscillators are restricted to their lowest Fock states, the annihilation operator can be mapped to the lowering operator of spin-1/2 thus yielding the Eq. (\ref{Cij}) that are suitable for our model \cite{NajmehPRR2020Appendix}. The modulus of $C_{jl}(t)$ indicates the degree of correlation between two systems. In particular $|C_{jl}|=1$ implies the two system are completely correlated and the time-evolution of the observables are phase-locking, while $|C_{jl}|=0$ means those two operators are completely uncorrelated.

As the counterpart of the classical synchronization, we put the investigation of the quantum synchronization in the framework of stochastic quantum trajectory. The idea of quantum trajectory method is that, under the diffusion limit, the time-evolution of the state of the system is represented by a set of individual quantum trajectories and the Lindblad master equation is shown to be equivalent to the following stochastic Schr{\"o}dinger equation \cite{breuer_book},
\begin{equation}
\text{d}|\psi(t)\rangle=D^{1}[|\psi(t)\rangle]dt +\sum_{k} D^{2}_{k}[|\psi(t)\rangle]dW_{k}(t),
\label{SSE}
\end{equation}
with the drift term
\begin{equation}
D^{1}[|\psi(t)\rangle] = [-i \hat{H}_{\mathrm{eff}}+\sum_{k} \frac{s_{k}(t)}{2}(\hat{\mathcal{O}}_{k}-\frac{s_{k}(t)}{4})]|\psi(t)\rangle,
\label{SSEPart1}
\end{equation}
and the diffusion term
\begin{equation}
D^{2}_{k}[|\psi(t)\rangle] = (\hat{\mathcal{O}}_{k}-\frac{s_{k}(t)}{2})|\psi(t)\rangle,\
\label{SSEPart2}
\end{equation}
where $\hat{H}_{\mathrm{eff}} = \hat{H} - i\sum_{k}\hat{\mathcal{O}}_{k}^{\dagger}\hat{\mathcal{O}}_{k}/2$ is the effective Hamiltonian and the scalar quantity $s_{k}(t) = \langle \psi(t)| (\hat{\mathcal{O}}_{k} + \hat{\mathcal{O}}_{k}^{\dagger})|\psi(t)\rangle$ is the expectation value of a linear combination of the dissipators. The random variable $dW_{k}(t)$ with the standard normal distribution is the stochastic Wiener increment obeying the It$\bar{\text{o}}$ rule $dW_{k}^{2} = dt$.

\section{Results}
\label{results}

\subsection{The Liouvillian spectrum}

\begin{figure}[!htpb]
\includegraphics[width=0.5\textwidth]{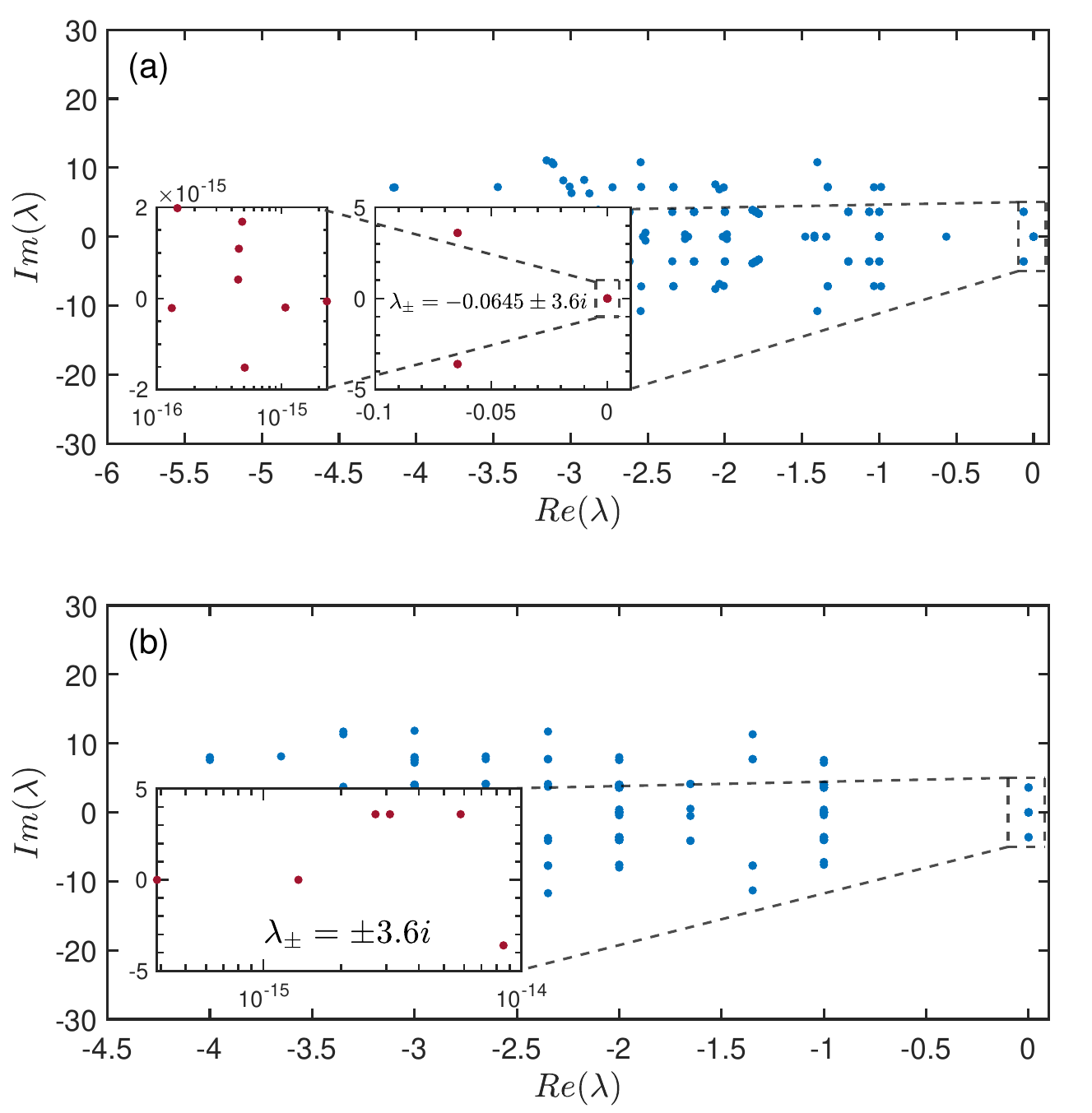}
\caption{\label{Spectrum} The Liouvillian spectra in the complex plane for the dissipative XYZ (a) and XXZ (b) models. The parameters are given as $\{J_{x},J_y,J_z\}/\Gamma = \{0.8,1,0.9\}$ for panel (a) and $\{J_{x},J_y,J_z\}/\Gamma= \{1,1,0.9\}$ for panel (b). Each inset is a detailed display of the eigenvalues near the zero value of the real part in the complex plane. In panel (a), the degeneracy of the zero eigenvalue is $D_n=10$ while in panel (b) the inset shows the existence of the degenerated zero eigenvalue and a pair of conjugated purely imaginary eigenvalue within the machine error.}
\end{figure}

As a preliminary exploration, we investigate the Liouvillian spectra of the XYZ and XXZ models. The eigenvalues of the different Liouvillians on the complex plane are shown in Fig. \ref{Spectrum}. In Fig. \ref{Spectrum}(a), one can find that the Liouvillian spectrum of the dissipative XYZ model is degenerated. More details are shown in the insets of Fig. \ref{Spectrum}(a) in which the spectra near the origin are zoomed in. The degeneracy of the zero eigenvalue is $D_{n}=10$ (not all the zero eigenvalues are visible in the inset due to the overlap) and the algebraic multiplicity is identical to the geometric one. The degenerate zero eigenvalues remind us that the system has multiple steady states, although the final steady state of the system depends on the initial state in a realistic evolution \cite{MingantiPRA2018}.
As shown in the right inset of Fig. \ref{Spectrum}(a), a pair of conjugate complex eigenvalues $\lambda{\pm}$, which have the largest (negative) real parts, are present. As mentioned in Sec. \ref{Model and Method}, the real parts of $\lambda_{\pm}$ are the so-called asymptotic decay rate  meaning that the system will evolve asymptotically into a time-independent steady state with the asymptotic decay rate being $\gamma=-0.0645$.

Despite sharing the identical dissipators with the XYZ model, the Liouvillian spectrum of the XXZ model shows relatively non-trivial results. Seeing from the fact that the number of the eigenvalues is much less than $\dim(\mathcal{L}) = 2^8$ in Fig. \ref{Spectrum}(b), the XXZ Liouvillian spectra is highly degenerated. Again we are interested in the eigenvalues with the real parts close to zero which are highlighted in the inset of Fig. \ref{Spectrum}(b). One can find that, in addition to the (degenerated) zero eigenvalues, there exists purely imaginary eigenvalues which can be regarded as the steady-state local phases in Eq. (\ref{time-evolution equation}). Such local phases do not vanish in the long-time limit so that the system will reach a non-stationary state in terms of persistent oscillation. Recall the steady state in Eq. (\ref{time-evolution equation}),
we can obtain the time-dependent expectation value of a local observable $\hat{o}$ as follows,
\begin{equation}
\lim_{t\to\infty}\langle\hat{o}\rangle_{t}=\sum_{\lambda^{*}_{j}=-\lambda_{j}}c_{j}e^{\lambda_{j}t}\text{Tr}[\hat{\rho}_{j}\hat{o}] + \text{const.}.
\label{eq:observables}
\end{equation}
Apparently, the period of the time-dependent $\langle\hat{o}\rangle_{t}$ is related to the purely imaginary eigenvalues. For instance, as will be seen in the next section, the highlighted purely imaginary eigenvalues $\lambda_{\pm}=\pm3.6i$ in the zoom-in of Fig. \ref{Spectrum}(b) results in a long-time oscillation with frequency $f=|\pm3.6|/2\pi\approx0.5730$.

We finish the discussion on the Liouvillian spectrum by noticing that if there are multiple purely imaginary eigenvalues, they should be commensurate (that is they are integer multiple of some fixed value) to guarantee the long-time periodic oscillation. Otherwise the time-evolution of the observable will either approach to an asymptotic stationary value or enter into chaotic region in the long-time limit \cite{Buca2022Sci}.

\subsection{The symmetry of the steady state}
\label{The symmetry of the steady state}
We notice that our system has a $\mathbb{Z}_{2}$-symmetry along the $z$-axis meaning that the collective operator $\hat{S}_{z}=\sum_{j}\hat{\sigma}^{z}_{j}$ has the following relations,
\begin{equation}
[\hat{H},\hat{S}_{z}] = 0,\quad  [\hat{\mathcal{O}}_{k},\hat{S}_{z}] = B_{k}\hat{\mathcal{O}}_{k},
\end{equation}
where $B_{k}$ is a real constant and the symmetry superoperator yields
\begin{equation}
\mathcal{S}[\cdot]=e^{i\frac{\pi}{2}\hat{S}_{z}}\cdot e^{-i\frac{\pi}{2}\hat{S}_{z}}.
\label{symmetry_op}
\end{equation}

\begin{figure}[!htpb]
\includegraphics[width=0.5\textwidth]{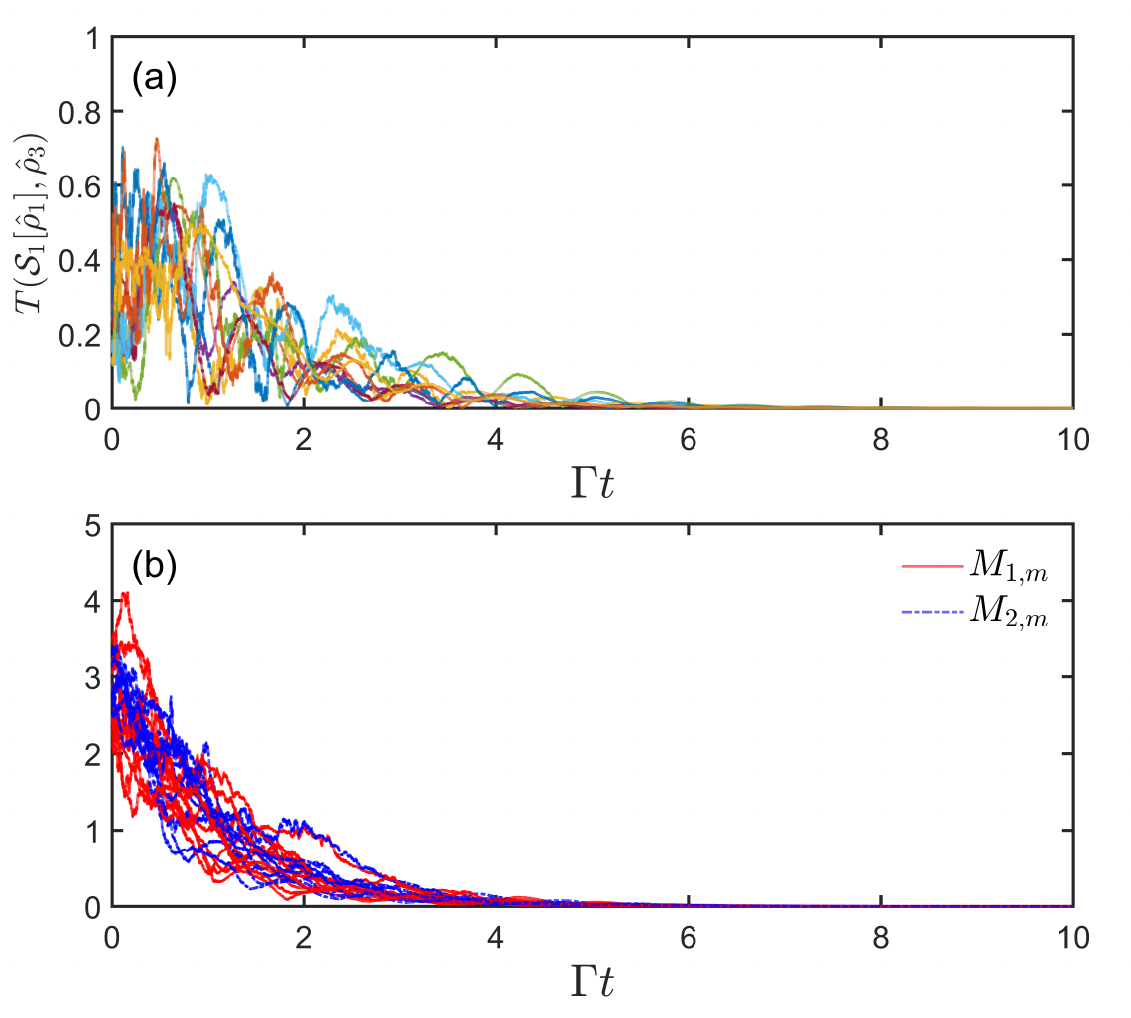}
\caption{\label{Mfig} (a) The time-dependence of the trace distance $T(\mathcal{S}_{1}[\hat{\rho}_{1}],\hat{\rho}_{3})$ (a) and  jump coefficient $M_{k,m}$ (b) for each single trajectory. The parameters are chosen as $\{J_{x},J_y,J_z\}/\Gamma=\{1,1,0.9\}$ and the time interval is set to $\text{d}t=10^{-3}\Gamma^{-1}$. }
\end{figure}

We further investigate the relationship between the states of (next-nearest neighboring) spins $1$ and $3$, $\hat{\rho}_{1}(t)$ and $\hat{\rho}_{3}(t)$, during the time-evolution. Here $\hat{\rho}_j(t)=\text{Tr}_{\neq j}[\hat{\rho(t)}]$ is the reduced density matrix of the $j$-th spin which is obtained by taking the partial trace of the density matrix of the total system.
We implement the $\mathcal{S}[\cdot]$ transformation on spin $1$ and show the trace distance between the transformed state $\mathcal{S}[\hat{\rho}_1(t)]$ and $\hat{\rho}_3(t)$ in Fig. \ref{Mfig}(a). The trace distance characterizes the similarity between two density matrices, say $\hat{\eta}_{1}$ and $\hat{\eta}_{2}$, and is given by
\begin{equation}
T(\hat{\eta}_{1},\hat{\eta}_{2})=\frac{1}{2}\text{Tr}\left[\sqrt{(\hat{\eta}_{1}-\hat{\eta}_{2})^{\dagger}(\hat{\eta}_{1}-\hat{\eta}_{2}) } \right],
\end{equation}
where $0\leq T(\hat{\eta}_{1},\hat{\eta}_{2})\leq 1$ and $T(\hat{\eta}_{1},\hat{\eta}_{2})=0$ for two indistinguishable states.
One can see that, starting with random initial state, the trace distance $T(\mathcal{S}_{1}[\hat{\rho}_{1}],\hat{\rho}_{3})$ evolve to zero after $\Gamma t\gtrsim8$ for all the trajectories. Since the transformation $\mathcal{S}$ represents a $\pi$-rotation along the $z$-axis, i.e. $\{\sigma^x,\sigma^y,\sigma^z\}\rightarrow\{-\sigma^x,-\sigma^y,\sigma^z\}$, the vanishing $T(\mathcal{S}_{1}[\hat{\rho}_{1}],\hat{\rho}_{3})$ means the magnetizations of spin $1$ and $3$ are always aligned oppositely in the $x$-$y$ plane during the long time-evolution. The magnetizations of spins $2$ and $4$ evolve in the same manner. Moreover one concludes that the system does not have spontaneous total magnetization in the $x$-$y$ plane. The scalar quantity $s_{k}(t) = \langle \psi(t)| (\hat{\mathcal{O}}_{k} + \hat{\mathcal{O}}_{k}^{\dagger})|\psi(t)\rangle$ in Eq. (\ref{SSE}) will be zero and thus will not make any contribution to the stochastic fluctuations in quantum trajectories.

We further discuss the reasons for the suppression of the stochastic fluctuation in the long-time evolution of each single trajectory. In principle, as described in Eq. (\ref{SSE}), it is the coupling to the environment that leads to such stochastic fluctuation during the time-evolution. Thus the disappearance of the fluctuation implies the decoupling of the environment. To corroborate, we define the following jump coefficient $M_{k,m}$
\begin{equation}
M_{k,m}=\big|\hat{\mathcal{O}}_{k}|\psi_{m}\rangle\big|.
\end{equation}
which is the modulus of the vector obtained by the jump operator $\hat{\mathcal{O}}_{k}$ acting on the state vector. The coefficient $M_{k,m}$ captures the effects of the action of the $k$-th jump operator on the $m$-th quantum trajectory. When the coefficient $M_{k,m}$ vanishes, it means that action of the jump operator on the wave function produces a null vector, namely, the coupling between the system and environment is switched off.

We show the time-dependence of $M_{k,m}$ in Fig. \ref{Mfig}(b) for more details. One can see that for any non-local jump operator, after a short time interval ($\Gamma t\leq6$), the jump coefficients $M_{k,m}$ rapidly approach to zero for all the trajectories. So the terms of $\hat{\mathcal{O}}_{k}|\psi_{m}\rangle$ as well as the scalar quantity $s_{k}(t) = \langle \psi(t)| (\hat{\mathcal{O}}_{k} + \hat{\mathcal{O}}_{k}^{\dagger})|\psi(t)\rangle$ become zero and will not contribute in the SSE (\ref{SSE}). The evolution of the trajectory is governed by the Hamiltonian.

Here we emphasize that not all the initial states will lead to stable oscillation in the long-time dynamics. For instance, we can define an invariant as follows
\begin{equation}
[\hat{H},\hat{G}]=[\hat{\mathcal{O}}_{k},\hat{G}]=[\hat{\mathcal{O}}_{k}^\dagger,\hat{G}]=0, j=1,2,	
\label{strongsymmetry}
\end{equation}
where $\hat{G}=e^{i\pi\sum_{\langle j,l\rangle,\alpha}\hat{\sigma}^{\alpha}_{j}\hat{\sigma}^{\alpha}_{l}/6}+\text{h.c.}$ ($\alpha=x,y,z$) is Hermitian (the choice of $\hat{G}$ may not be unique). So there is at least one eigenstate of $\hat{G}$ is a simultaneous dark states of the Hamiltonian and jump operator and remains unchanged during the evolution. More generally, Eq. (\ref{strongsymmetry}) defines a `strong symmetry' identified by $\hat{G}$ which allows the simultaneous block-decomposition of the Hamiltonian and the jump operators \cite{tindall2023}. Therefore if the initial state is a superposition of
the states belonging to different symmetry subspaces, i.e. the state space spanned by the degenerated eigenvectors, the state of each individual trajectory will evolve into one of the superposed subspaces at random and then be frozen there.

\subsection{The oscillations in long-time dynamics}
\begin{figure*}[!htpb]
\includegraphics[width=1\textwidth]{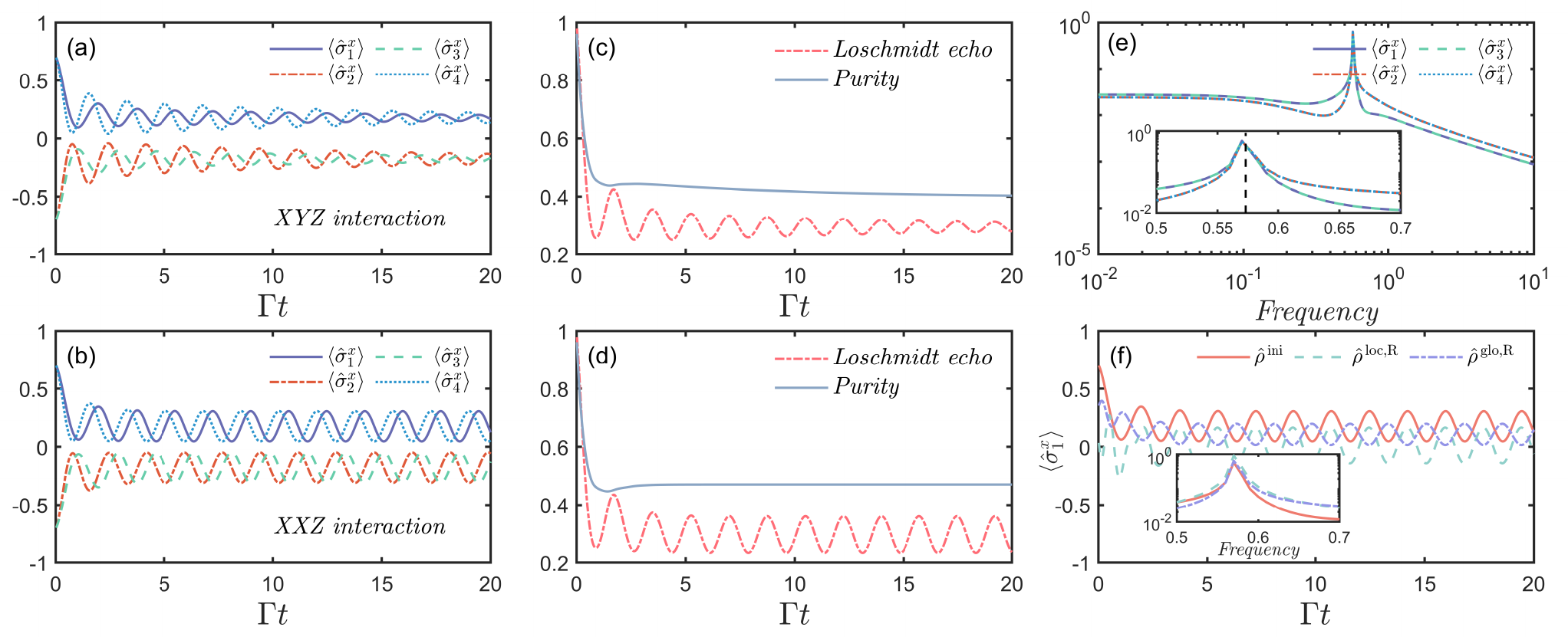}
\caption{\label{dynamics} The expectation values of the local observables $\langle\hat{\sigma}_j^x\rangle$ ($j=1,2,3$ and $4$) as a function of time for the XYZ (a) and XXZ (b) models. Panels (c) and (d) display the time-dependence of the Loschmidt echo (red dashed lines) and purity (blue solid lines) in accordance with panels (a) and (b), respectively. The initial states of both models are $\hat{\rho}^{\text{ini}}=|\psi^{\text{ini}}_{j}\rangle\langle\psi^{\text{ini}}_{j}|$. (e) The Fourier spectra of the time-dependent $\langle\hat{\sigma}^x_j\rangle$ presented in panel (b). The vertical dashed line in the inset denotes the frequency corresponding to the imaginary part of $\lambda_{\pm}$. (f) The time-evolution of $\langle\hat{\sigma}^{x}_{1}\rangle$ for the XXZ model with different initial states. The inset displays the corresponding Fourier spectra.
The parameters are chosen as $\{J_{x},J_y,J_z\}/\Gamma=\{0.8,1,0.9\}$ for the XYZ model and $\{J_{x},J_y,J_z\}/\Gamma=\{1,1,0.9\}$ for the XXZ model.}
\end{figure*}

The dynamics of the XYZ and XXZ model also differ due to the different Liouvillian spectra. In Fig. \ref{dynamics} we show the time-evolution of the expectation values of spin magetizations for both models. The initial states of each spin are chosen as the normalized states on the equatorial plane of the Bloch sphere, i.e. $|\psi^{\text{ini}}_{j}\rangle=(|\uparrow\rangle + e^{i\phi_{j}}|\downarrow\rangle)/\sqrt{2}$, with $\phi_{j}=(2j-1)\pi /4$,($j=1,2,3,4$) for concreteness. Therefore, the initial states for the joint system is given by $|\psi^{\text{ini}}\rangle=\bigotimes_{j=1}^{4}{|\psi^{\text{ini}}_{j}\rangle}$.

For the XYZ model one can see that at the early stage the $\langle\sigma_j^x\rangle$ oscillates with time, as shown in Fig. \ref{dynamics}(a). However, such early-stage oscillations is damping towards to an asymptotic steady state for sufficient long time. This is consistent with the property of the Liouvillian spectrum shown in Fig. \ref{Spectrum}(a) where the asymptotic decay rate $\gamma$ is nonzero (although small).

In Fig. \ref{dynamics}(b), we show the time-evolution of the same observables for the XXZ model. In contrast to the XYZ model, the magnetizations exhibit irregular evolution at the initial moment and soon enter into stable oscillations with a well defined period. These behaviors are again consistent with the Liouvillian spectrum shown in Fig. \ref{Spectrum}(b) where the purely imaginary eigenvalues appear.

In Figs. \ref{dynamics}(a) and (b), the time-dependence of the magnetizations of the spins sharing the common reservoir already shows negative correlation indicating that the spins $1$ and $3$ (or spins $2$ and $4$) are anti-synchronized. This can be witnessed by the so-called Pearson correlator as discussed in the Appendix. We will come back to the discussion of the synchronization in Sec. \ref{Quantum synchronization} in the sense of quantum trajectories.

Now we are going to investigate behaviors of the so-called Loschmidt echo $L(t)$ and the purity which are concerning in the experimental detection of the oscillatory dynamics. The Loschmidt echo incorporates the simple idea for indicating the sensitive changes in the system state after being affected by the perturbation. It has been used to probe for the quantum chaos \cite{GorinPR2006}, quantum time crystal \cite{BookerNJP2020}, dynamical phase transition \cite{Heyl2018RPP}, and information scrambling \cite{Yan2020PRL}. The Loschmidt echo is defined as
\begin{equation}
L(t) = \text{Tr}[\hat{\rho}^{\dagger}(t)\hat{\rho}(0)],
\label{L_echo}
\end{equation}
and quantifies the overlap between the initial state $\hat{\rho}(0)$ and the state $\hat{\rho}(t)$ at time $t$. As a signal for the periodic oscillation, the Loschmidt echo should be a persistent oscillation \cite{BookerNJP2020}.
In Figs. \ref{dynamics}(c) and (d), we show the time-dependence of the Loschmidt echo $L(t)$ for the XYZ and XXZ models, respectively. One can see that the amplitude of the oscillating Loschmidt echo shrinks as the time pasts in the XYZ model implying the absence of oscillations in the long-time limit. While for the XXZ model the Loschmidt echo oscillates with constant amplitude and a well-defined period signaling the stable oscillation of the magnetization.

The purity of the system state during the evolution is shown in Figs. \ref{dynamics}(c) and \ref{dynamics}(d). The purity is defined by $P(t)=\text{Tr}[\hat{\rho}^2(t)]$. One can see that, different from the dynamical behaviors of Loschmidt echo, the purity does not oscillate during the time-evolution for both the models. It is interesting that, as shown in Fig. \ref{dynamics}(d) for the XXZ model, the purity remains unchanged as soon as the stable oscillation of the magnetization is established.

In order to extract the frequency of the oscillation, we perform the fast Fourier transformation to convert the time-dependent magnetizations shown in Fig. \ref{dynamics}(b) in the frequency domain. As shown in Fig. \ref{dynamics}(e) each curve exhibits the peak around an identical dominant frequency $f_d\approx0.57$ which is consistent with the Liouvillian spectrum analysis, i.e. $2\pi f_d\sim \text{Im}[\lambda_\pm]$ with $\lambda_\pm=\pm 3.6i$ being the purely imaginary eigenvalues in Fig. \ref{Spectrum}(b).

Furthermore, we focus on the dynamical results of the XXZ model with different initial states in Fig. \ref{dynamics}(f). The random product and correlated states are set as follows:

{\it The product random state} $\hat{\rho}^{\text{loc, R}}$. Each spin is initialized as a normalized random state $|\psi^{\text{R}}_{j}\rangle$ with $(j=1,2,3,4)$, then we can generate the initial state $\hat{\rho}^{\text{loc, R}}=|\psi^{\text{loc,R}}\rangle\langle\psi^{\text{loc,R}}|$ by the product state $|\psi^{\text{loc,R}}\rangle=\bigotimes_{j=1}^{4}|\psi^{\text{R}}_{j}\rangle$;

{\it The correlated random state} $\hat{\rho}^{\text{glo, R}}$. This initial state can be directly generated by the global (normalized) random state $\hat{\rho}^{\text{glo, R}}=|\psi^{\text{glo, R}}\rangle\langle\psi^{\text{glo, R}}|$.

As shown in Fig. \ref{dynamics}(f), despite the speed of reaching stable oscillations and the amplitudes of oscillations being different, all those three cases show similar periodic behaviors. We also perform the FFT analysis on these data and present the results in a zoomed-in inset of Fig.\ref{dynamics}(f). The numerics indicate that the three initial states produce the oscillations with the identical frequency. This once again confirms that the oscillatory behavior of the system originates from the particular structure of the master equation.

\subsection{The largest Lyapunov exponent}
Here we unravel the quantum master equation (\ref{Masterequation}) into the stochastic Schr{\"o}dinger equation (\ref{SSE}) and primarily focus on the time-evolution of single quantum trajectory of the system. These trajectories can be considered as the quantum analogy of the classical trajectories in the phase space. The initial states of all the trajectories $|\psi_{m}\rangle$ are random states. We find that the long-time oscillation can be observed in each individual trajectory. The stability of the oscillation can be verified by the so-called largest Lyapunov exponent during the time-evolution as the following \cite{YusipovChaos2019,XingliPRB2021,YusipovChaos2021,YusipovChaos2022}.

In analogy to the classical definition, the quantum largest Lyapunov exponent $\lambda$ characterizes the average growth \jj{rate} of the ``distance'' between two initial states which are close to each other. The distance is determined through the expectation
value of an observable, for instance as will be seen soon we use the magnetization of one of the spins $\hat{\sigma}^x_1$. It has been shown that the values of the Lyapunov exponent do not depend much on a particular choice of the observable \cite{YusipovChaos2022}. In a practice realization, it is convenient to extract the quantum Lyapunov exponent by monitoring the evolution of the distance between a single trajectory, also notated as the fiducial trajectory, and an auxiliary trajectory. We denote the initial state of the fiducial trajectory by $|\psi_{f}(0)\rangle$, while the state of the auxiliary trajectory is initialized by $|\psi_{a}(0)\rangle$ = ($|\psi_{f}(0)\rangle + \delta|\psi\rangle)/|||\psi_{f}(0)\rangle + |\delta\psi\rangle)||$ where $|\delta\psi\rangle$ is the perturbation.

When the time-evolution is launched, the expectation values of the magnetizations with respect to the states of the auxiliary and fiducial trajectories, $\langle \hat{\sigma}^{x}_{1}(t)\rangle_a$ and $\langle \hat{\sigma}^{x}_{1}(t)\rangle_f$,  starts to deviate. We denote the ``distance" of these two trajectories by the difference $\Delta(t)=\langle \hat{\sigma}^{x}_{1}(t)\rangle_a-\langle \hat{\sigma}^{x}_{1}(t)\rangle_f$ and the initial distance as $\Delta_0=\Delta(0)$. As the time pasts, the distance may diverge in an exponential way $\Delta(t)\sim\Delta_0 e^{\lambda t}$. Here we empirically set a threshold $\Delta_{\text{max}}$ of the distance. If the distance between the two trajectories evolves beyond such threshold $|\Delta(t_0)|>\Delta_{\text{max}}$, signaling a possible chaotic behavior for the first time, at some point of time $t_0$, the growth factor $d(t_0) =|\Delta(t_0)|/\Delta_{0}$ is recorded. Since the quantum Lyapunov exponent is determined by the averaged growth factor within a single trajectory (the fiducial trajectory), the monitoring starts over by resetting the state of the auxiliary trajectory to be a perturbed state of the fiducial trajectory $|\psi(t_0)\rangle_a=|\psi(t_0)\rangle_f+|\delta\psi\rangle$ and record the growth factor $d(t_k) =|\Delta(t_k)|/\Delta_{0}$ if $|\Delta(t_k)|>\Delta_{\text{max}}$ at any future time $t_k$ ($k>0$). Such process is repeated up to a sufficient long time $t\rightarrow\infty$, the largest Lyapunov exponent is thus obtained by averaging all the growth factors,
\begin{equation}
\lambda = \lim_{t\to\infty}\frac{1}{t}\sum_{k}\ln{d_{k}(t_{k})}.
\label{LyapunovExponent}
\end{equation}

\begin{figure}[!htp]
\includegraphics[width=0.5\textwidth]{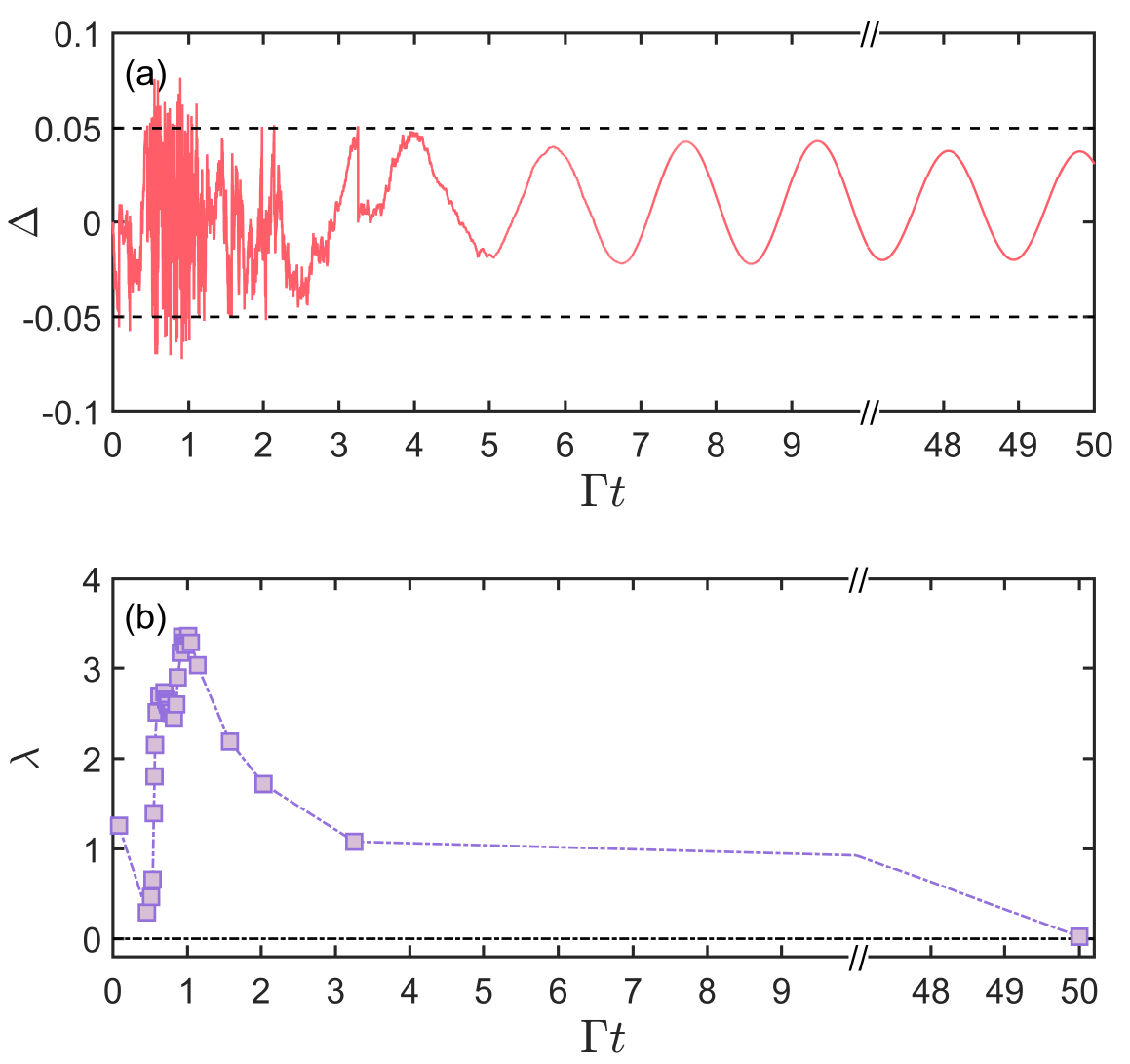}
\caption{\label{Lyapunov}(a) The difference of observables $\Delta(t)$ varies with time (red line) and the black dashed lines denote threshold $\Delta_{\text{max}} = 0.05$. (b) The largest Lyapunov exponent changes in time. When the absolute value of $\Delta(t)$ exceeds the threshold $\Delta_{\text{max}}$, an update of the largest Lyapunov exponent takes place. }
\end{figure}

Although the instantaneous quantum Lyapunov exponent may be nonzero due to the stochastic fluctuation,  $\lambda$ will approach to zero in the limit of $t\rightarrow\infty$ for stable oscillation. In Fig. \ref{Lyapunov}(a) one can see that at the early stage of the time-evolution, the distance $\Delta(t)$ between the fiducial and auxiliary trajectories shows rapid and drastic variation and goes beyond the threshold $\Delta_{\text{max}}$ (shown as the black dash lines) multiple times. After multiple-time resetting of the state of the auxiliary trajectory, the ``distance'' trends to be stable and stays below the threshold. The process of updating the largest Lyapunov exponent over time is shown in Fig.\ref{Lyapunov}(b). Corresponding to the results in Fig.\ref{Lyapunov}(a), the largest Lyapunov exponent is updated multiple times iteratively at the initial moment. After experiencing unstable evolution in a short period, the largest Lyapunov exponent decreases monotonically. In the long-time evolution ($\Gamma t=50$ in the current simulation), the value of $\lambda$ eventually descends to zero hinting that the system is robustness to the perturbation and the oscillations are stable.

\subsection{Quantum synchronization}
\label{Quantum synchronization}
\begin{figure}[!htpb]
\includegraphics[width=0.5\textwidth]{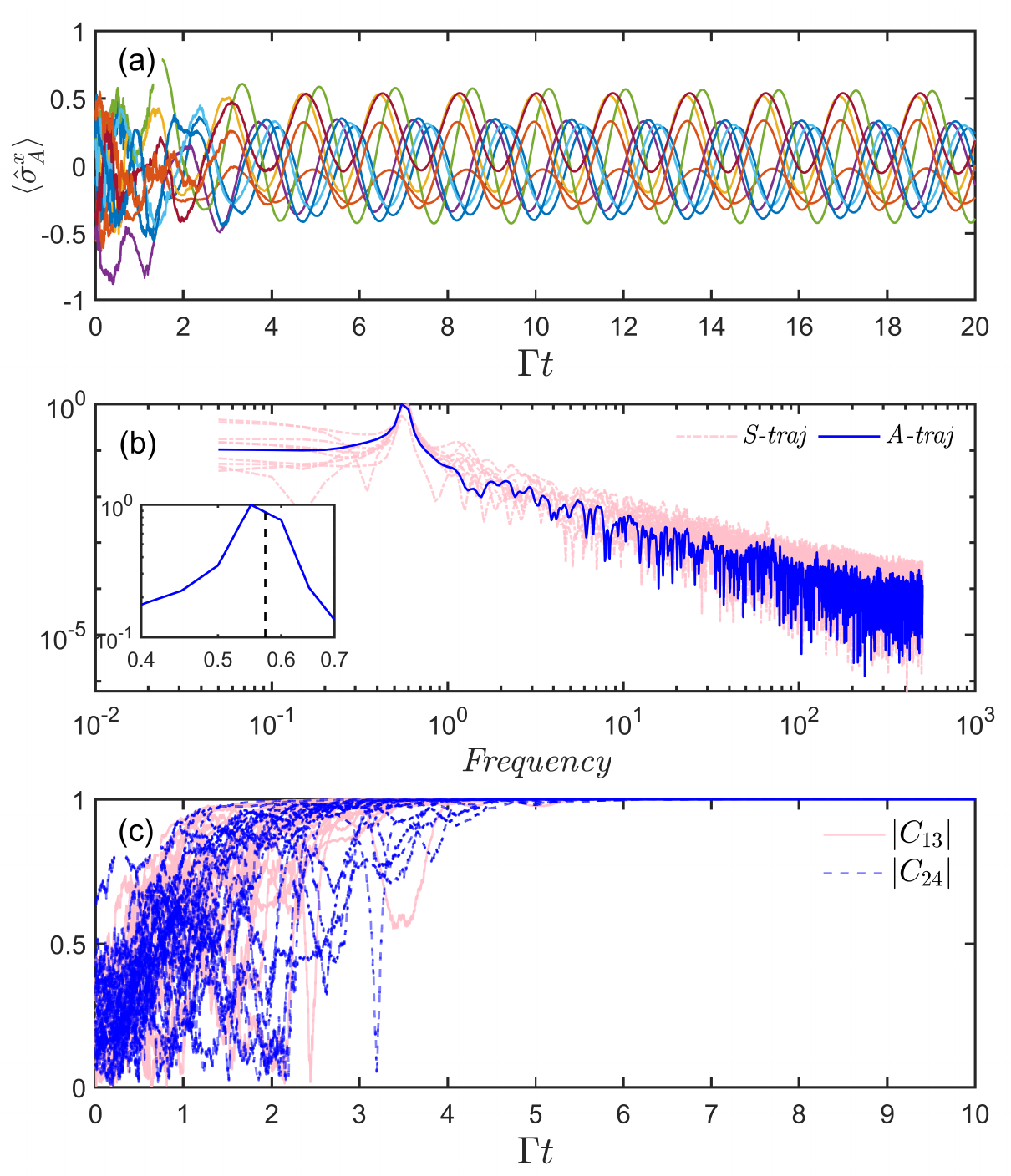}
\caption{\label{Trajectory} (a) The expectation values of the observable $\langle\hat{\sigma}^{x}_{1}\rangle$ obtained by SSE vary with time. The lines with different colors indicate the results of $n_{\text{traj}}=10$ randomly chosen quantum trajectories. The total number of quantum trajectories is $N_{\text{traj}}=100$. (b) The Fourier spectrum for the set of single quantum trajectory {\it(S-traj)} presented in panel (a) and for the average quantum trajectory {\it(A-traj)}. The details around the peak of the average trajectory is shown in the inset. (c) The corresponding time-evolution of the modulus of $|C_{jk}|$ for the set of single quantum trajectories presented in panel (a). The parameters are chosen as $\{J_{x},J_y,J_z\}/\Gamma=\{1,1,0.9\}$ and the time interval is $\text{d}t=10^{-3}\Gamma^{-1}$. }
\end{figure}

So far we have confirmed the existence of the persistent oscillation in the dissipative XXZ model. Now we are in a position to investigate the synchronization among the oscillations of the magnetizations of different spins. In practise, we simulated with a total of $N_{\text{traj}}=100$ quantum trajectories and randomly selected $n_{\text{traj}}=10$ trajectories to show the results in Fig. \ref{Trajectory}.

As shown in Fig. \ref{Trajectory}(a) for the time-evolution of $\langle\hat{\sigma}^{x}_{1}\rangle$ of the $10$ randomly selected quantum trajectories, one can see that at the initial stage, i.e. $0\leq \Gamma t\leq 6$, the time-dependence of $\langle\hat{\sigma}^{x}_{1}\rangle$ for all the $10$ selected trajectories show completely stochastic behaviors. As the evolution continues, the randomness (fluctuations) of the quantum trajectories are gradually suppressed, instead the curves of each $\langle\hat{\sigma}^{x}_{1}\rangle$ become smooth.

In the general stochastic evolutions in quantum systems, the fluctuations in each trajectory can be observed through the whole time-evolution as reported in Refs. \cite{NajmehPRR2020,JinPRX2016,RotaNJP2018,JinPRB2018}. However, it is interesting that in our system the stable oscillation is established in each individual trajectory and the single quantum trajectory is not affected by stochastic processes in the long-time evolution. We will explain this phenomenon in Sec.\ref{The symmetry of the steady state} in the context of properties such as the symmetry of the system.

In Fig. \ref{Trajectory}(b), we perform the FFT on the time-dependence of $\langle\sigma^{x}_{1}\rangle$ for the selected trajectories and on the  averaged $\langle\sigma^{x}_{1}\rangle$ over all the $N_{\text{traj}}=100$ trajectories, respectively. The Fourier spectra show that the peaks for all the single trajectories appear (almost) at an identical dominant frequency. In the zoom-in of Fig. \ref{Trajectory}(b), we compare the Fourier spectrum of the averaged $\langle\sigma^{x}_{1}\rangle$ and the dominant frequency $f_d\approx0.5730$ obtained from the Liouvillian spectra analysis. The result shows good agreement meaning that each of the single trajectory oscillates individually with the same frequency $f_d$ and differs only in the amplitudes. We note that the peaks for some of the single trajectories are not prominent because the stable oscillations have not been established yet in the present time window (up to $\Gamma t=20$).

In order to characterize the synchronization among the oscillations of the spins, we employ the correlator in Eq. (\ref{Cij}) to quantify the degree of synchronization of the quantum stochastic trajectories of two spins \cite{NajmehPRR2020}.
In Fig. \ref{Trajectory}(c), we show the feature of synchronization between next-nearest neighboring spins for a set of $10$ selected trajectories. By substituting the corresponding operators $\{\sigma_1^+,\sigma_3^-\}$ and $\{\sigma_2^+,\sigma_4^-\}$ into Eq. (\ref{Cij}) we find that in all the trajectories the next-nearest neighboring spins are completely correlated indicated by $|C_{13}|$ and $|C_{24}|$ approaching to unity for $\Gamma t\gtrsim 8$. This means that there is always a phase-locking between the spins.

\subsection{Infinite-size spin lattice}
In recent years, searching for the persistent, rigid and periodic oscillation in the steady state of a macroscopic dissipative system (in the thermodynamics limit) has attracted considerable attention. Such the oscillation in the long-time dynamics signals the breaking of the time-translation symmetry in the system and it is intimately related to the time-crystalline phase \cite{Iemini2018PRL,Tuquero2022PRA,Taheri2022NC,Prazeres2021PRB,LledNJP2020,Hajdusek2022PRL}.

In particular, the existence of oscillatory phase in the infinite-size dissipative spin-1/2 with nearest-neighboring XYZ interaction, the same as Eq. (\ref{Hamiltonian}), has been discussed \cite{Chan2015PRA}.
It is shown that the long-time oscillations do not exist in the system with only local jump operators $\hat{\sigma}^-_m$, $\forall m$, i.e. each site is subjected to an individual decay \cite{owen2018}. However, when the next-nearest-neighboring (n.n.n.) interactions is present, the oscillatory phase emerges due to the frustration in the system \cite{XingliPRB2021}. The frustration introduced by the coherent interactions between next-nearest-neighboring sites is responsible for the appearance of the long-time oscillations. This motivates us to check the possibility of emergence of long-time oscillations in the presence of the n.n.n. incoherent dissipation, i.e. with the nonlocal jump operators $\sqrt{\Gamma}(\hat{\sigma}^-_m+\hat{\sigma}^-_n)$  with $m$ and $n$ being n.n.n. sites. To this aim, we extend our model, described by Eqs. (\ref{Hamiltonian}) and (\ref{Masterequation}), into infinite square lattice to have a glance on the effect of nonlocal dissipation.

However solving the master equation of dissipative quantum many-body system is rather demanding since the dimension of the state space grows exponentially as the size of the system increasing.
In the past decades, although several methods have been developed, such as cluster mean-field approximation \cite{JinPRX2016}, linked-cluster expansion method \cite{BiellaPRB2018}, corner-space renormalization \cite{FinazziPRL2015,RotaPRL2019}, and machine learning techniques \cite{YoshiokaPRB2019,NagyPRL2019,HartmannPRL2019,VicentiniPRL2019} (a comprehensive review on the numerical methods can be found in Ref. \cite{weimer2021}), it is still challenging to simulate the dynamics of open quantum many-body system.

\begin{figure}[!htp]
\includegraphics[width=0.5\textwidth]{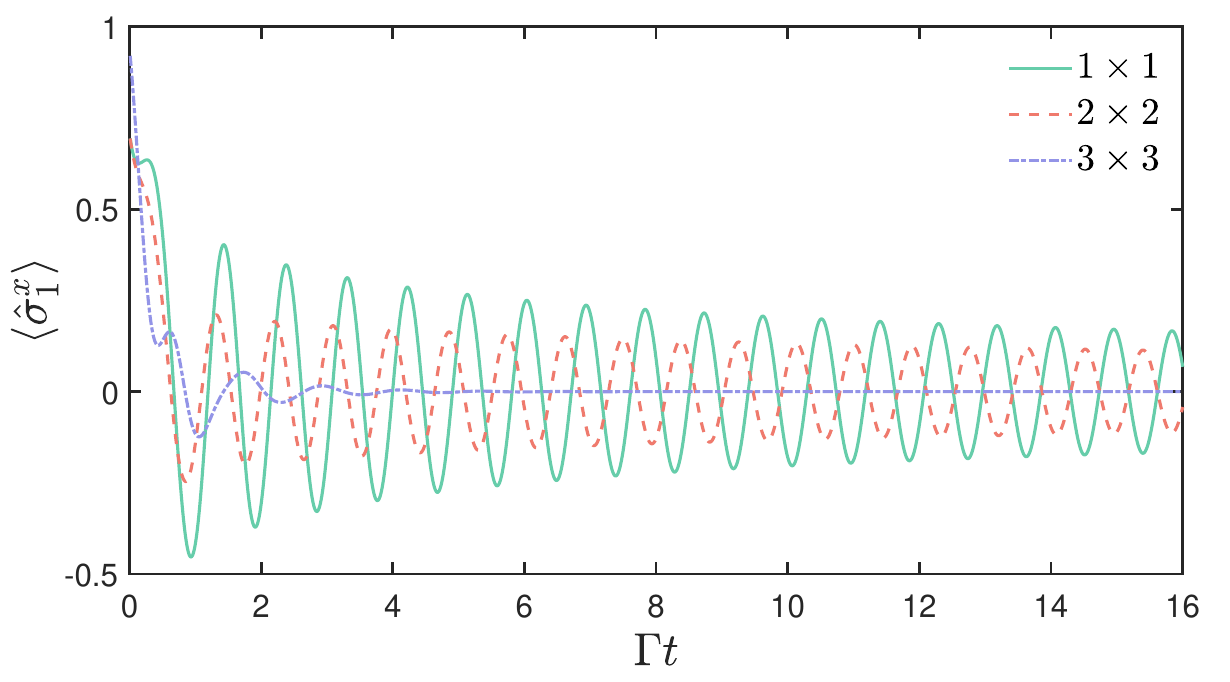}
\caption{\label{CMF} The cluster mean-field time-evolution of the magnetization $\langle\hat{\sigma}^{x}_{1}\rangle$ for $1\times1$ (standard), $2\times2$ and $3\times3$ clusters. The initial state of each state is initialized on the equatorial plane of the Bloch sphere. The parameters are chosen as $\{J_{x},J_y,J_z\}/\Gamma=\{1,1,0.9\}$.}
\end{figure}

In this work, we implement the cluster mean-field (CMF) approximation on the system to explore its steady-state properties in the thermodynamic limit. The basic idea of the CMF approximation is to factorize state of the many-body system $\hat{\rho}$ into the tensor product of the state $\hat{\rho}_{\mathcal{C}}$ of a set of identical clusters through the factorization ansatz, i.e., $\hat{\rho}=\bigotimes_{C}\hat{\rho}_{\mathcal{C}}$ ($\hat{\rho}_C=\hat{\rho}_{C'}$, $\forall C, C'$). With the help of the CMF approximation, the master equation for the state of the infinite-size system can be decoupled into the following equation,
\begin{equation}
\label{CMF_ME}
\frac{\text{d}\hat{\rho}_C}{\text{d}t}=-i\left[\hat{H}_{C}+\hat{H}_{C-C'}+\hat{H}_{L},\hat{\rho}_C\right]+\mathcal{D}_{C}[\hat{\rho}_C]+\mathcal{D}_{L}[\hat{\rho}_C],
\end{equation}
where $\hat{H}_{C}=\sum_{\alpha,\langle m,m'\rangle\in C}{J_\alpha\hat{\sigma}^\alpha_m\hat{\sigma}^\alpha_{m'}}$ describes the on-cluster interactions between nearest-neighboring sites $\langle m,m'\rangle$ of cluster $C$, $\hat{H}_{C-C'}=\sum_{\alpha,m\in\partial{C}, m'\in \partial{C'}}{J_\alpha\langle\hat{\sigma}_{m'}^\alpha\rangle\hat{\sigma}_m^\alpha}$ is the mean-field Hamiltonian induced by the inter-cluster interactions between the nearest neighboring sites on the boundaries $\partial{C}$ and $\partial{C'}$ of cluster $C$ and its nearest-neighbor, respectively, $\hat{H}_{L}=\frac{\Gamma}{2}\sum_{m\in\partial{C}, m''\in\partial{C'}}{(-i\langle\hat{\sigma}_{m''}^+\rangle\hat{\sigma}_m^-+\text{h.c.})}$ is the effective local Hamiltonian induced by the nonlocal jump operators associated to the next-nearest-neighboring pairs $\langle m,m''\rangle$ belonging to all the nearest-neighboring clusters $C'$, and $\alpha=x,y,z$ denote the components of the Pauli matrices. In addition, the dissipator $\mathcal{D}_{C}[\hat{\rho}_C]=\sum_{\langle m,m'\rangle\in C}{(\hat{\mathcal{O}}_{mm'}\hat{\rho}_C\hat{\mathcal{O}}_{mm'}^\dagger-\frac{1}{2}\{\hat{\mathcal{O}}_{mm'}^\dagger\hat{\mathcal{O}}_{mm'},\hat{\rho}_C\})}$ describes the on-cluster dissipation with the non-local jump operators $\hat{\mathcal{O}}_{mm'}=\sqrt{\Gamma}(\hat{\sigma}^-_m+\hat{\sigma}^-_{m'})$ and the dissipator $\mathcal{D}_{L}[\hat{\rho}_C]=\sum_{m\in\partial{C},m''\in\partial{C}' }{(\hat{\mathcal{O}}_m\hat{\rho}_C\hat{\mathcal{O}}_m^\dagger-\frac{1}{2}\{\hat{\mathcal{O}}_m^\dagger\hat{\mathcal{O}}_m,\hat{\rho}_C\})}$ describes the effective dissipation induced by the CMF factorization on the boundary of $C$ with the local jump operators being $\hat{\mathcal{O}}_m=\sqrt{\Gamma}\hat{\sigma}^-_m$. When considering the single-site cluster, the CMF analysis is reduced to the standard mean-field treatment as all the correlations in the lattice are neglected. In principle, as the size of cluster increasing the short-range correlations are gradually included and thus the property of the system in the thermodynamic limit is obtained. More details about the CMF approximation can be found in Ref. \cite{JinPRX2016}.

In Fig. \ref{CMF}, we show the time-evolution of the magnetization $\langle\hat{\sigma}^{x}_{1}\rangle$ of the dissipative XXZ model for various sizes of clusters. The state of each site in a chosen cluster is initialized in the equatorial plane of the Bloch sphere. Although the oscillatory behaviors are observed in the levels of $1\times1$ and $2\times2$ clusters, the system relaxes to asymptotic steady state up to $3\times3$ cluster which excludes the existence of the time-translation breaking phase in the steady state of the infinite system within the chosen parameters.

Although our result rules out the oscillatory phase for the chosen parameters in the infinite-size XXZ model with nonlocal dissipation, there are still interesting problem open. For example, in the all-to-all XXZ model, the oscillatory phase is shown to appear in the collective decay case while disappear in the local decay case \cite{lee2014pra}. The former case indicates a infinite long-range correlated dissipation while the latter indicates a local dissipation. How does the oscillatory phase emerge as the range of correlated jump operator increase and whether there is a critical length of the correlated jump operator would be interesting for future studies. Along this line, the inclusion of the n.n.n. nonlocal jump operator is an expedient point to start with.

\section{Summary}
\label{Summary}
In summary, we have investigated the phenomenon of synchronization in spin systems with non-local dissipation. In the absence of the external driving, the system with XXZ interaction generates the oscillations in the long-time dynamics due to the non-local jump operators. This is manifested by the existence of the purely imaginary eigenvalues of the Liouvillian superoperator associated to the Lindblad quantum master equation.

We have unravelled the quantum master equation into quantum trajectories whose time-evolution are effectively described by the SSE. It is interesting that the long-time periodic oscillation can be observed even in each single trajectory. The stability of the single-trajectory oscillation has been verified by the converging to zero of the largest Lyapunov exponent during the time-evolution. The periodicity of the oscillation extracted by the FFT is in consistent with the one predicted by the purely imaginary eigenvalues of the Liouvillian. In particular, in contrast to the results reported by previous literatures, the random fluctuations during the time-evolution caused by the coupling to the environment is high suppressed in our considered model. This can be attributed to the decoupling of the system from the environment by means of the jump operator becomes zero as soon as the stable oscillation is established.

We have also investigated the synchronization among the oscillation of each spin. By adopting the measure of synchronization proposed in Ref.  \cite{NajmehPRR2020}, we have found that the next-nearest-neighboring spins is completely synchronized indicating a phase-locking between the spins during the time-evolution. We have also revealed the sublattice symmetry in our model through investigating the time-dependence of the trace distance between the states of a given spin and its next-nearest neighbor.

The oscillatory behavior in the long-time dynamics of our considered few-body model is due to the action of the non-local jump operators generates a subspace which is the eigenspace of the system Hamiltonian but is dark to the jump operators. Although the CMF analysis on the infinite-size system governed by the quantum master equation (\ref{CMF_ME}) excludes the existence of the time-translation symmetry breaking phase with the chosen parameter, whether the non-local dissipations in the interacting spin system may induce long-time oscillation is still open. Finally, we believe that it will be interesting to look for the oscillatory or time-crystalline phases in open quantum many-body system with the appropriately designed non-local dissipations.

\section*{ACKNOWLEDGMENTS}
This work is supported by National Natural Science Foundation of China under Grant No. 11975064.

\section*{APPENDIX}
\label{appendix}
The Pearson correlation coefficient is a measure of the strength of a linear association dependence of two variables. It takes a range of values from $-1$ to $+1$. A value of $0$ indicates that the two variables are uncorrelated. A value greater (less) than zero indicates a positvie (negative) association between the two variables, that is as the value of one variable increases, the value of the other variable increases (decreases). Specially, a value of $+1$ or $-1$ means that the two variables are in total positive or negative correlation.

\begin{figure}[!htp]
\includegraphics[width=0.5\textwidth]{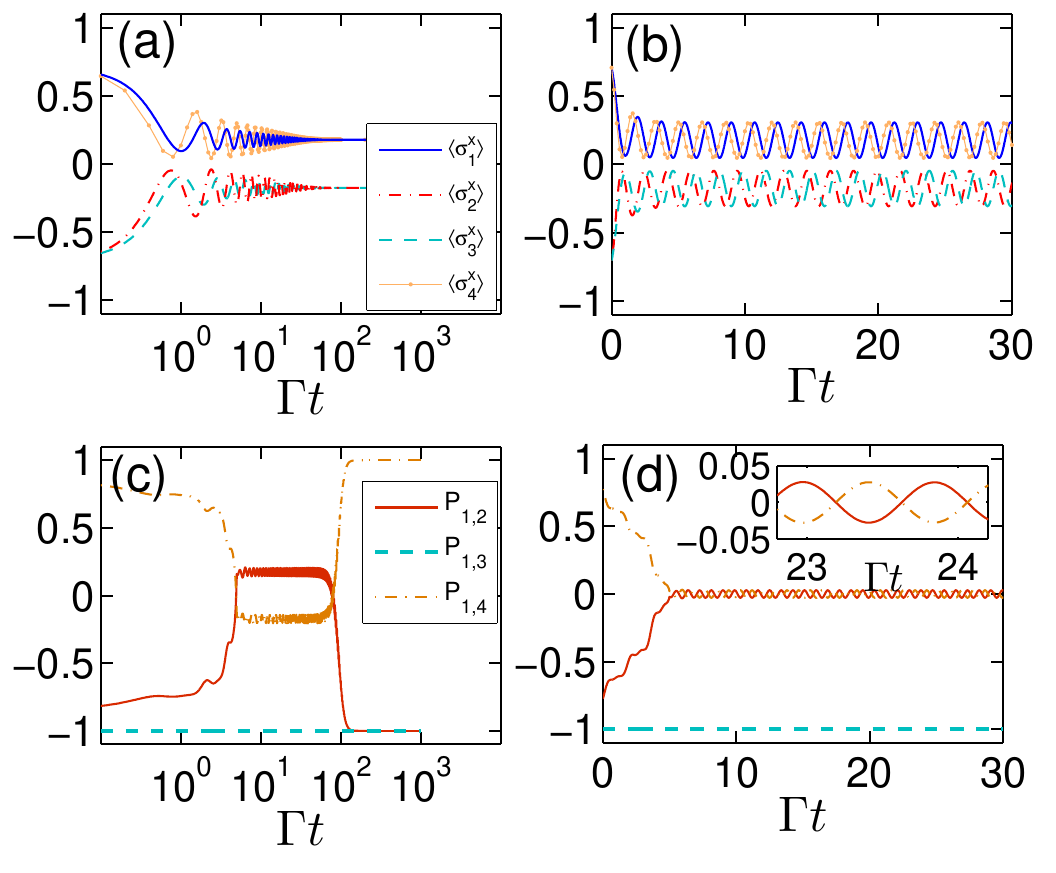}
\caption{\label{PearsonCo} The time evolution of the magnetization $\langle\hat{\sigma}^x_i\rangle$ in the XYZ model (a) and XXZ model (b). The Pearson correlation coefficient between spin 1 and other spins in the XYZ model (c) and XXZ model (d). The inset in (d) is a zoom in for $P_{1,2}$ and $P_{1,4}$. The width of time window in producing the Pearson correlation coefficient is $\Delta t=10\Gamma^{-1}$.  Other parameters are chosen as $\{J_{x},J_y,J_z\}/\Gamma=\{0.8,1,0.9\}$ for the XYZ model and $\{J_{x},J_y,J_z\}/\Gamma=\{1,1,0.9\}$ for the XXZ model. Note that the time-axis for the XYZ model is in the logarithmic scale.}
\end{figure}

Considering two functions in time, the Pearson correlation coefficient can be used to measure the temporal correlation between them. Recently, the Pearson correlation coefficient has been employed to witnessing the synchronization emerging in the dynamics of both classical and quantum systems \cite{Karpat2019PRA,NajmehPRR2020}. Given two time-dependent expectation values of local observables $\hat{\sigma}_1$ and $\hat{\sigma}_2$,  for example, the magnetizations of two different spins in the models discussed in the main text, the Pearson coefficients is defined as \cite{GalvearXiv},
\begin{equation}
P_{\hat{\sigma}_1,\hat{\sigma}_2}(t|\Delta t)=\frac{\int^{t+\Delta t/2}_{t-\Delta t/2}{(\hat{\sigma}_1-\bar{\hat{\sigma}}_1)(\hat{\sigma}_2-\bar{\hat{\sigma}}_2)\text{d}t}}{\sqrt{\int^{t+\Delta t/2}_{t-\Delta t/2}{(\hat{\sigma}_1-\bar{\hat{\sigma}}_1)^2\text{d}t}\int^{t+\Delta t/2}_{t-\Delta t/2}{(\hat{\sigma}_2-\bar{\hat{\sigma}}_2)^2\text{d}t}}},
\end{equation}
where
\begin{equation}
\bar{\hat{\sigma}}_i=\frac{1}{\Delta t}\int^{t+\frac{\Delta t}{2}}_{t-\frac{\Delta t}{2}}{\hat{\sigma}_i(t)\text{d}t},
\end{equation}
and $\Delta t$ the length of a sliding window. We apply the Pearson coefficient to the time-dependent magnetization in Figs. \ref{dynamics}(a) and (b) and show the results in Fig. \ref{PearsonCo}. We have prolonged the time axis to show a full view of the dynamics.

From Fig. \ref{PearsonCo}(a) one can see that the expectation values of the magnetizations in XYZ model eventually approach to asymptotic steady-state values due to the non-zero real parts of the Liouvillian eigenvalues. As shown in Fig. \ref{PearsonCo}(c),  the early-stage time-evolution of the magnetizations of spins $1$ and $3$, which share a common reservoir, are negative correlated with $P_{1,3}=-1$. While the Person coefficients $P_{1,2}$ and $P_{1,4}$ describing the correlations between th time-dependent $\langle\hat{\sigma}^x_1\rangle$ with $\langle\hat{\sigma}^x_2\rangle$ and $\langle\hat{\sigma}^x_4\rangle$, respectively, shows nonzero values before the oscillations are damped. Due the fact that spins $2$ and $4$ are in negative correlation, $P_{1,2}$ and $P_{1,4}$ take different signs. As the time pasts, the system evolves into time-independent steady-state ($\Gamma t\gtrsim 10^2$) the Pearson coefficients become meaningless because all the magnetizations are constant.

In Fig. \ref{PearsonCo}(b), we show the time-evolution of the magnetizations of each spin in the XXZ model. The oscillations survive in the long-time limit because of the existence of purely imaginary Liouvillian eigenvalues. Again, one can see that the time-dependent magnetizations of spins $1$ and $3$ are always negative correlated, $P_{1,3}=-1$, as show in Fig. \ref{PearsonCo}(b). This is in consistence with the conclusion from the analysis with the complex-valued correlator (\ref{Cij}). The Pearson coefficient $P_{1,2}$ and $P_{1,4}$ oscillate around zero periodically because there exists a stable phase difference between time-evolutions of $\langle\hat{\sigma}^x_1$ and $\langle\hat{\sigma}^x_2\rangle$ ($\langle\hat{\sigma}^x_4\rangle$). The details of $P_{1,2}$ and $P_{1,4}$ are shown in the inset of Fig. \ref{PearsonCo}(d). Again $P_{1,2}$ and $P_{1,4}$ always take different signs because the spins $2$ and $4$ are negative correlated.

\end{document}